\documentclass[12pt,preprint]{aastex}
\usepackage{natbib}
\begin{document}

\title{On the Initial Conditions for Star Formation and the IMF}

\author{Bruce G. Elmegreen}
\affil{IBM Research Division, T.J. Watson Research Center, 1101
Kitchawan Road, Yorktown Heights, NY 10598} \email{bge@watson.ibm.com}

\begin{abstract}
Density probability distribution functions (PDFs) for turbulent
self-gravitating clouds should be convolutions of the local log-normal
PDF, which depends on the local average density $\rho_{\rm ave}$ and
Mach number ${\cal M}$, and the probability distribution functions for
$\rho_{\rm ave}$ and ${\cal M}$, which depend on the overall cloud
structure. When self-gravity drives a cloud to increased central
density, the total PDF develops an extended tail. If there is a
critical density or column density for star formation, then the
fraction of the local mass exceeding this threshold becomes higher near
the cloud center. These elements of cloud structure should be in place
before significant star formation begins. Then the efficiency is high
so that bound clusters form rapidly, and the stellar initial mass
function (IMF) has an imprint in the gas before destructive radiation
from young stars can erase it. The IMF could arise from a power-law
distribution of mass for cloud structure. These structures should form
stars down to the thermal Jeans mass $M_J$ at each density in excess of
a threshold. The high-density tail of the PDF, combined with additional
fragmentation in each star-forming core, extends the IMF into the Brown
Dwarf regime. The core fragmentation process is distinct from the cloud
structuring process and introduces an independent {\it core
fragmentation mass function} (CFMF). The CFMF would show up primarily
below the IMF peak.
\end{abstract}

\section{Introduction}

The density probability distribution function for supersonically
turbulent gas is a log-normal when self-gravity is not important
\citep{vazquez94,pvs98,pnj97,lemaster08,federrath10,price11}. A power
law extension at high density appears with self-gravity
\citep{klessen00,vs08,kritsuk10}. Observations of column density show
extended PDFs in star-forming regions and nearly pure log-normal PDFs
in non-star-forming regions, which is consistent with these theoretical
trends \citep{kainulainen09,froebrich10,lombardi10}.

Here we consider a log-normal PDF that applies locally for the local
average density and Mach number in a cloud. The PDF for the whole cloud
is the convolution of this with the probability distribution function
for these quantities, which vary with position when gravity and energy
dissipation are important. The result has a power-law tail if the Mach
number is constant, and a gradually falling tail if the Mach number
decreases with increasing average density.

The convolution-PDF depends on the ratio of maximum to minimum average
cloud density, i.e., the core-to-edge average density ratio in a GMC.
The PDF also depends on cloud mass because the maximum Mach number
does. As a cloud evolves from a diffuse state to a strongly
self-gravitating state, the density PDF develops an extended tail where
the star formation rate and the efficiency per unit free fall time
become high.

The convolution-PDF is a useful starting point for approximations of
internal cloud structure. We use it to determine several interesting
characteristics of star formation: the cumulative mass as a function of
density, the fraction of the local mass in the form of high-density
cores, and the efficiency of star formation per unit free fall time,
all as functions of average density or cloud radius. The first of these
derived quantities illustrates why the gas consumption time for most
tracers is longer than the dynamical time. The second and third are
important for understanding why bound clusters form.

Cloud structure may also be relevant to the stellar initial mass
function \citep[IMF; see review in][]{shad11}.  Pre-stellar gas cores
share many properties with protostars and young stars, including the
mass distribution function in the power-law regime
\citep[e.g.,][]{motte98,rathborne09}, the autocorrelated positions of
each \citep{johnstone00,johnstone01,enoch06,young06,schmeja08}, and the
mass fractions compared to the whole cloud.  Here we consider the IMF
model by \cite{padoan02} with some modifications to determine IMFs in
clouds with convolution-PDFs.

Cloud structure characterized approximately by a convolution-PDF could
be in place before significant star formation begins. This would help
explain observations of short formation times and high cluster
efficiencies, even when the efficiency per unit free fall time is small
on average \citep{krum07}. It might also explain the persistence of a
near-universal IMF in the presence of intense stellar feedback. While
some radiative heating may be important to limit the overproduction of
Brown Dwarfs \citep{bate09}, intense radiation from massive stars can
prevent the formation or collapse of low-mass cores, giving a top-heavy
IMF \citep{krum10} unlike that usually observed in regions of massive
star formation.

In what follows, we present convolution-PDFs and derived quantities for
various ratios of center-to-edge average cloud densities and Mach
numbers (Sect. 2). Applications to the IMF are discussed in Section 3.

\section{Convolution PDFs}

The density PDF in a region of the interstellar medium depends on the
distribution of local Mach number ${\cal M}$ and local average density,
$\rho_{ave}$, and may be written
\begin{equation}
P_{\rm PDF}(\rho)=\int_{\rho_{\rm ave\;min}}^{\rho_{\rm ave\;max}}
P_{\rm PDF,local}(\rho | \rho_{\rm ave})P_{\rm ave}(\rho_{\rm ave})
d\rho_{\rm ave}.\label{eq1}
\end{equation}
$P_{\rm PDF,local}(\rho | \rho_{\rm ave})$ is the conditional
probability distribution function for density $\rho$, given the average
$\rho_{\rm ave}$. We use the usual log-normal relation for this PDF,
assuming it comes from supersonic turbulence:
\begin{equation}P_{\rm PDF,local}=(2\pi D^2)^{-1/2}e^{-0.5\left(\ln(
\rho/\rho_{\rm pk})/D\right)^2}
\end{equation}
per unit $\ln(\rho)$. The density $\rho_{\rm pk}$ is at the peak of the
local PDF and the log-normal width is $D$.  The peak and average
densities are related by
\begin{equation}
\rho_{\rm pk}=\rho_{\rm ave}e^{-0.5D^2},\end{equation} and the Mach
number and log-normal width are related approximately by \citep{pnj97}
\begin{equation}
D^2=\ln(1+0.25{\cal M}^2).\label{eq:D}\end{equation}

The probability distribution function for the average density, $P_{\rm
ave}(\rho_{\rm ave})$, and the relation between Mach number and
density, depend on the history of the cloud.  A cloud that has just
formed from some gas collection process or a cloud that is
gravitationally unbound will be somewhat uniform in average density,
while a cloud that has contracted gravitationally or is collapsing will
be centrally condensed. To cover this range of conditions, we assume
that the average density varies with radius in a spherical cloud as
\begin{equation}
\rho_{\rm ave}(r)=\rho_{\rm edge} {{ r_{\rm edge}^\alpha + r_{\rm
core}^\alpha}\over {r^\alpha + r_{\rm
core}^\alpha}}\label{eq:rho}\end{equation} where the degree of central
condensation depends on the ratio
\begin{equation}{\cal C}={{\rho_{\rm ave}(r=0)}\over{\rho_{\rm edge}}}.
\end{equation}
The core radius, $r_{\rm core}$, is used to avoid a density singularity
and to control the maximum density contrast in the average cloud. The
probability distribution function for $\rho_{\rm ave}$ is given by the
relative volumes of different densities,
\begin{equation}
P(\rho_{\rm ave})=4\pi r^2 \left(d\rho_{\rm ave}/dr\right)^{-1}.
\end{equation}
After normalization and the definition $z=\rho_{\rm edge}/\rho_{\rm
ave}(r)$, with $r_{\rm core}/r_{\rm edge}=\left({\cal
C}-1\right)^{-1/\alpha}$, we get
\begin{equation}
P(\rho_{\rm ave})d\rho_{\rm ave}={{3{\cal C}\left(z{\cal
C}-1\right)^{(3-\alpha)/\alpha}}\over{\alpha \left({\cal
C}-1\right)^{3/\alpha}}}dz.\end{equation}

The resultant PDF for all density is then, from equation (\ref{eq1})
\begin{equation}
P_{\rm PDF,total}(y)= {{3{\cal
C}}\over{\alpha(2\pi)^{0.5}}}\int_{1/{\cal C}}^{1}
\exp\left(-{{\ln^2(yze^{0.5D^2})}\over{2D^2}}\right) {{\left(z{\cal
C}-1\right)^{(3-\alpha)/\alpha}}\over{D\left({\cal
C}-1\right)^{3/\alpha}}}dz,\label{eq:pdftotal}
\end{equation}
per unit $\ln y$, where we have normalized the local density to the
average cloud edge density, $y=\rho/\rho_{\rm edge}$. Note that $y$ is
the local normalized density, including turbulent fluctuations, while
$z$ is the inverse of the average density, not including the turbulent
fluctuations, both normalized to the edge density. The width $D$ of the
log-normal generally varies with position $r$, and so $D$ can be a
function of $z$.

For large $yz$, the exponent becomes small and the fluctuations
infrequent.  Thus the integral is dominated by $z<1/y$ at large $y$,
and for constant $D$, scales with only the density distribution for
average density, which means $P_{\rm PDF,total}\propto y^{-3/\alpha}$.
This is the power law part of the density PDF at large density relative
to the cloud edge \citep{kritsuk10}.

\begin{figure}\epsscale{.9}
\plotone{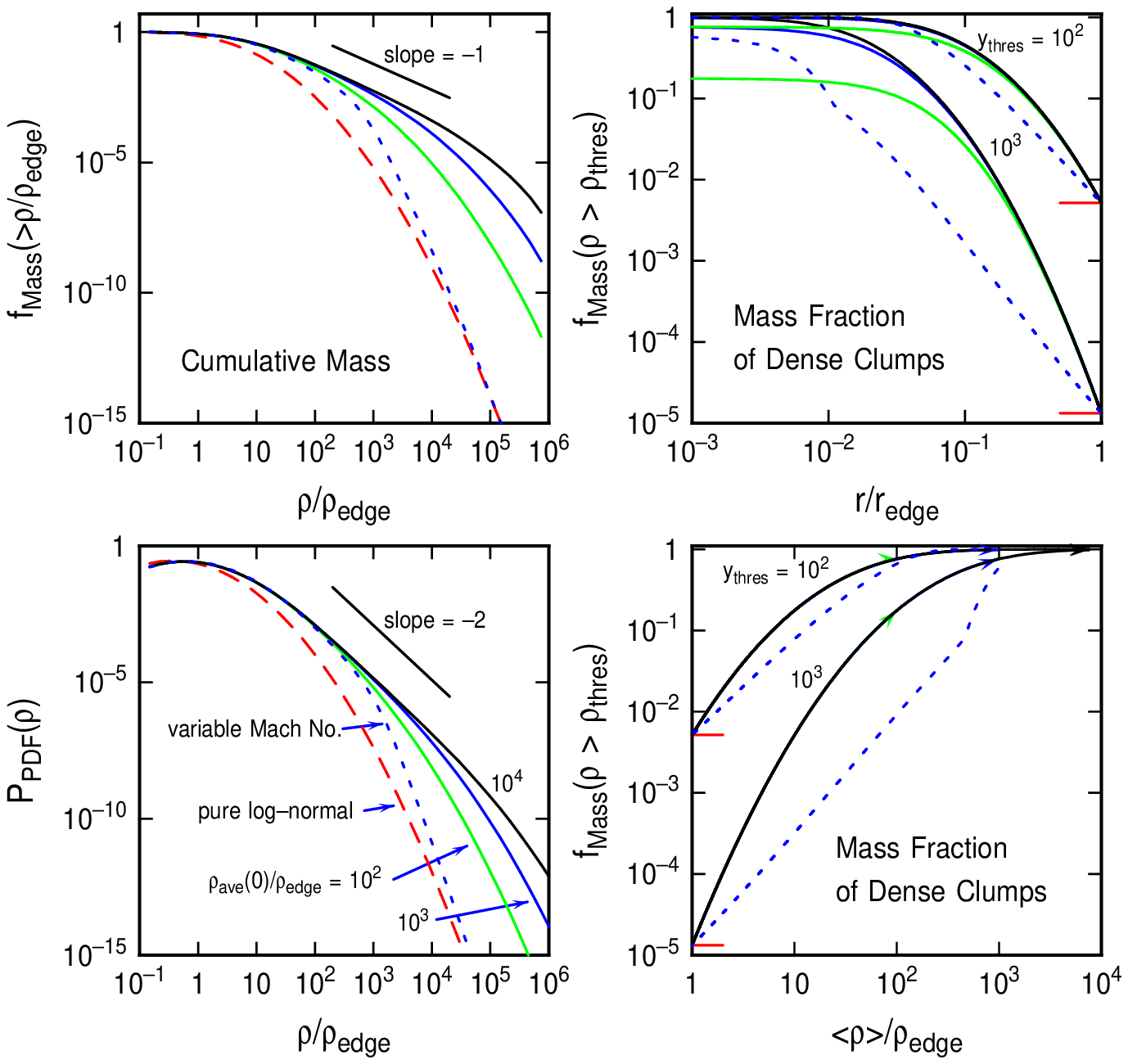} \caption{(Lower left) The convolution-PDF for local
density in a cloud is shown as a solid curve for 3 values of the
concentration ${\cal C}=\rho_{\rm ave}(0)/\rho_{\rm edge}$. The PDF
develops a power-law tail with slope $3/\alpha$ for high central
concentrations in a cloud with a density profile given by eq.
(\ref{eq:rho}). The dashed red line is a pure log-normal PDF with the
same dispersion $D$ as the convolution PDF. The blue dotted line is a
convolution-PDF with a variable $D$ in the log-normal part, as given by
eqs. (\ref{eq:D}) and (\ref{eq:mach}). (Top Left) The cumulative mass
function for the same PDFs as in the bottom left, according to eq.
\ref{eq:fmass}. (Right panels) The fraction of the cloud mass at a
density greater than the threshold density $y_{\rm thres}$ versus
normalized average density (bottom) and normalized cloud radius (top).
Two values of $y_{\rm thres}$ are shown, and for each, three values of
${\cal C}$ are shown (${\cal C}=10^2$, $10^3$, and $10^4$ for green,
blue, and black curves). In the bottom panel, the 3 ${\cal C}'s$
overlap and their endpoints are shown by appropriately colored arrows.
The dotted blue curves on the right have variable $D$ with ${\cal
C}=10^3$ and with $y_{\rm thres}=10^2$ and $10^3$, as evident from
their respective proximities to the solid curves.}
\label{convolution}\end{figure}

Figure \ref{convolution} shows the total density PDF in the lower left
panel for three ratios of the core-to-edge average density contrast,
${\cal C}$, in the case of constant Mach number equal to 5 (i.e.,
constant $D=1.41$) and radial-density slope $\alpha=1.5$.  These are
the three solid-line curves in the figure. The dashed red curve is a
pure log-normal PDF with the same constant $D$.  The abscissa is the
normalized local density, not the average, and so is not a simple
function of radius. In fact, each local density $\rho$ corresponds to a
wide range of possible radii depending on the local compressions (i.e.,
on the local log-normal PDFs). The figure indicates that the PDF
develops a power-law tail as ${\cal C}$ increases. The slope of this
power-law is indicated by the straight line segment, which has a slope
of $-3/\alpha=-2$.  For larger ${\cal C}$, the length of the power law
segment increases.

Figure \ref{convolution} also shows a case with variable Mach number
${\cal M}$ as a dotted blue line. We consider here a cloud with
virialized sub-condensations, in which case the local Mach number
scales with $\left(M[r]/r\right)^{1/2}$. This is normalized to some
Mach number at the cloud edge, ${\cal M}_{\rm edge}$:
\begin{equation}
{\cal M}(r) = {\cal M}_{\rm edge}\left(M[r]r_{\rm edge}/M_{\rm total}
r\right)^{1/2}.\label{eq:mach}\end{equation} The radial mass variation
is from $M(r)=\int_0^r 4\pi r^2\rho_{\rm ave}(r) dr$.  For an
approximately uniform column density, equation (\ref{eq:mach}) gives
the \cite{larson81} size-linewidth law. We consider a minimum ${\cal
M}=1$ to get density fluctuations from turbulence. Thus ${\cal M}$ is
taken to be the larger of 1 and the value given by equation
(\ref{eq:mach}). When ${\cal M}$ varies in this way, it gets smaller
for higher average density and closer to the cloud center. Then the
log-normal Kernel of the convolution-PDF gets narrower with increasing
$\rho_{\rm ave}$, and the total PDF curves down below a power-law tail.
The power law tail occurs when the Mach number is constant throughout a
cloud. This situation may not be realistic for a cloud more than
several crossing times old, and if a constant ${\cal M}$ is observed,
it may indicate extreme cloud youth.

The upper left panel of Figure \ref{convolution} shows the cumulative
mass fractions for the same cases as in the bottom left panel. These
are the mass fractions for relative densities greater than the values
on the abscissa.  They were determined by integrating over all radii
for each relative local density, $y$, i.e., using equation
\ref{eq:pdftotal},
\begin{equation}
f_{\rm Mass}(>y)={{\int_y^{y_{\rm max}} P_{\rm PDF,total}(y)
dy}\over{\int_0^{y_{\rm max}} P_{\rm PDF,total}(y)  dy}}.
\label{eq:fmass}\end{equation} Note that $P_{\rm PDF,total}$ is defined
as per unit $\ln(y)$, so in fact we integrated over $y d\ln(y)$ for
equal intervals of $d\ln(y)$, but this is the same as $dy$ written
above. We assumed $y_{\rm max}=10^6$ for computational reasons. Figure
\ref{convolution} indicates that the mass fraction of dense gas at high
density is several orders of magnitude larger for a power-law tail than
for a log-normal PDF with the same Mach number. This means that with a
threshold density for star formation, the star formation rate should be
much larger in clouds that are centrally condensed (${\cal C}>>1$) than
in uniform clouds (${\cal C}=1$). This is in agreement with
observations \citep[e.g.,][]{lada09,lada10} and recent simulations
\citep{cho11}.

The bottom right panel of Figure \ref{convolution} shows the fraction
of the mass at a density larger than the threshold density, versus the
normalized average density. The top right panel shows the same quantity
as a function of normalized radius. This mass fraction of threshold gas
is not uniform throughout a cloud but increases as the average density
increases. If stars form in threshold gas, then they have a greater
probability per unit gas mass of forming near the center for ${\cal
C}>> 1$. This threshold gas fraction is also related to the efficiency
of star formation measured as a star-to-total mass ratio.  A high
threshold gas fraction favors the formation of bound clusters.

\begin{figure}\epsscale{.6}
\plotone{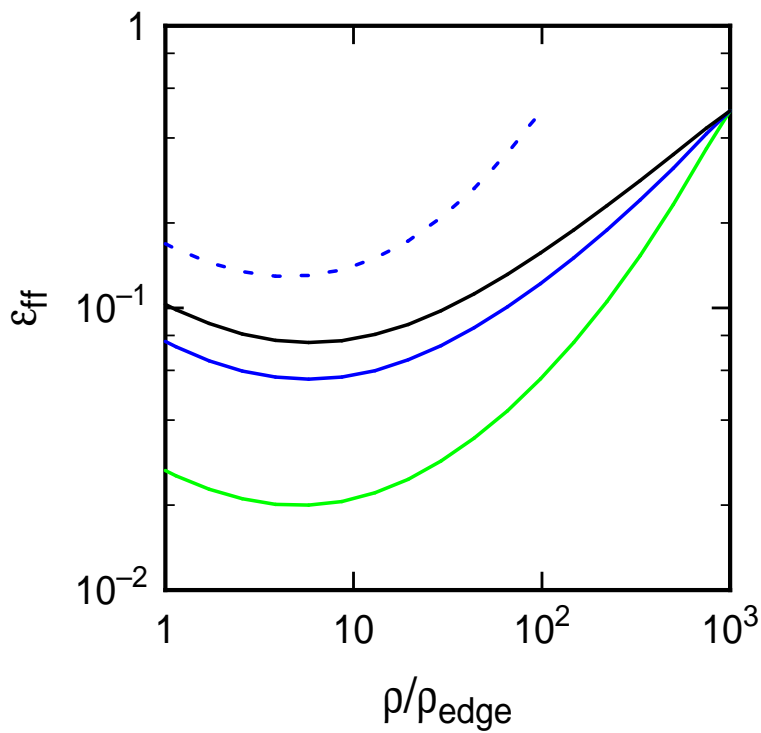} \caption{The local efficiency of star formation per
unit free fall time is plotted as a function of the local normalized
density $y=\rho/\rho_{\rm edge}$, from equation (\ref{eq:eps}). The
cloud concentration is ${\cal C}=10^2$, $10^3$, and $10^4$ for green,
blue, and black curves, and the threshold density is $y_{\rm
thres}=10^3$ with constant $D$. The dotted blue curve has variable $D$
with ${\cal C}=10^3$ and $y_{\rm
thres}=10^2$.}\label{convolution_eps}\end{figure}

Figure \ref{convolution_eps} shows the local efficiency of star
formation per unit dynamical time, $\epsilon_{\rm dyn}$, versus the
normalized local density.  The local efficiency is the fraction of the
gas mass that turns into stars in a local dynamical time, measured in
some part of the cloud where the density might be higher than average.
This concept is useful if stars form at a characteristic threshold
density, $\rho_{\rm thres}$, and evolution toward this density occurs
at a rate proportional to the local dynamical rate, $(G\rho)^{1/2}$. As
the local average density around a star-forming clump increases toward
the center of the clump, the fraction of the local mass at or above
this density that is also above the threshold density increases,
because the lower density gas that does not participate in star
formation gets left behind. This means $\epsilon(\rho)$ increases with
$\rho$. Numerical simulations use variable $\epsilon(\rho)$ also: if
the threshold for star formation in a simulation is taken to be a low
density, then the efficiency for star formation defined at that density
has to be low too, to get the overall rate correct
\citep[e.g.,][]{teyssier10}. If we consider a clumpy cloud with
contours at density $\rho$, then the average star formation rate per
unit volume inside these contours may be written
\begin{equation}{\rm SFR} = \epsilon_{\rm dyn}(\rho)(G\rho)^{1/2}
\int_\rho^\infty \rho P_{\rm PDF,tot}(\ln \rho) d\ln
\rho.\end{equation} The integral is the mass at densities larger than
$\rho$ inside the contours where $\epsilon_{\rm dyn}(\rho)$ is
measured; it also appears in equation (\ref{eq:fmass}).  Inside these
contours we can envision higher and higher densities until we reach the
threshold density, whereupon a fixed high fraction of the gas gets into
stars. We take this fixed fraction to be 0.5, in which case
\begin{equation}
\epsilon_{\rm dyn}(\rho)=0.5\left({{\rho_{\rm thres}}\over
{\rho}}\right)^{1/2} \left({{f_{\rm Mass}[\rho_{\rm thres}]} \over
{f_{\rm Mass}[\rho]}}\right). \label{eq:eps}
\end{equation}
This is the quantity shown in Figure \ref{convolution_eps}  using the
convolution PDF. A similar diagram for a log-normal PDF was shown in
\cite{e08}, where $\epsilon_{\rm dyn}(\rho)$ was considered in more
detail.

The importance of this result, and of the density dependence for total
efficiency (Fig. \ref{convolution} - right hand side), is that they
allow us to understand the formation of bound clusters in only a few
dynamical times, even though the average efficiency per unit free fall
time is observed to be low, like a few percent \citep[e.g.][]{krum07}.
The point is that a cloud does not have to wait for a time equal to the
dynamical time multiplied by the inverse of this average low efficiency
in order to build up a sufficiently high total efficiency to make a
bound cluster. This would require $\sim20$ dynamical times for the
average density, which is too long compared to observations
\citep{e07}. In fact, the efficiency per unit free fall time and the
total efficiency are high near the cloud center from the beginning of
the star formation process because most of the gas at $\rho_{\rm
thres}$ is initially clustered there.

\section{Applications to the IMF}

The density PDF may also have some relevance to the stellar IMF.
\cite{padoan02} considered the IMF to result from a power law function
of core mass $M$ multiplied by the probability that $M$ exceeds the
thermal Jeans mass $M_{\rm J}$. This probability equals the integral
over the distribution function of Jeans mass, from $M_{\rm J}=0$ to
$M_{\rm J}=M$. The distribution function for Jeans mass was related to
the density PDF in a one-to-one fashion using the relation between
$M_{\rm J}$ and density. For the density PDF, \cite{padoan02} assumed a
log-normal.  The extension of this log-normal into a power law PDF at
high density in the present paper affects their IMF at low mass because
the high density part contributes to the low $M_J$ values and to the
probability that $M>M_J$ at low $M_J$.

We follow this IMF model with some modifications. First, the mass
function for cloud structure in the supersonic regime is taken to be a
power law with a slope $\alpha$ dependent on the slope $\beta$ of the
density power spectrum, which for long lines of sight through a cloud,
is the same as the slope of the column density power spectrum. This
dependency is $\alpha\approx4.35-0.71\beta$ \citep{shad11}, and for
commonly observed $\beta=-2.8$, is $\alpha=2.35$, the Salpeter function
slope. Thus the mass function for stars is related to the mass function
for cloud structure in the supersonic regime, as in \cite{padoan02}.

These mass functions come primarily from the size distribution function
for substructure, which is a power law in turbulent media when the
power spectrum is a power law. Each mass equals approximately the
structure's volume multiplied by the boundary density used to define
that structure \citep{shad11}. As a result, the mass function for cloud
structure in the supersonic regime is nearly independent of the density
PDF. This model contrasts with that of \cite{henn08}, who assume that
the IMF comes directly from an integral over the PDF of log-density,
with stellar mass mapping out the Jeans mass so that low density
regions form high mass stars.  In the present model, all stars
throughout the power-law portion of the IMF form at all densities above
a threshold.

Second, gas forms stars only in regions that have a mass exceeding the
local thermal Jeans mass, regardless of the shape of these regions.
Elongated or irregular structures tend to collapse first into more
irregular shapes \citep[the eccentricity of an ellipsoidal shape
increases during collapse -- ][]{hunter62,fuji68}, and then gradually
into globules or cores that are collection points for gas draining down
filaments. The condition for strong self-gravity means that the
substructures defined by a certain density threshold, having a mass
larger than the thermal Jeans mass at that density, have a high
probability of turning into stars.

A third condition for star formation concerns the magnetic field.
Magnetic diffusion is not essential for star formation because the gas
can always move parallel to the field until the mass-to-flux ratio in
the collecting core exceeds the critical value for collapse
\citep[e.g.,][]{li10}. However, if the field energy density is large
compared to the gravitational energy density, then this motion has to
be over a long distance and that takes a lot of time. Turbulent media
mix gas on crossing time scales, so only processes that are either fast
or persistent can occur to completion. As a result, stars tend to form
where the field diffuses out rapidly.

This diffusion constraint means that star formation occurs in regions
that are shielded from background near-uv starlight by more than a few
magnitudes of visible-wavelength extinction. H$_2$ accumulation takes
only $\sim1$ magnitude of visual extinction \citep{spit75}, which is
not enough to prevent the ionization of elements heavier than Helium.
CO accumulation takes another magnitude or so of visible extinction,
depending on density. Neither condition alone is enough to cause the
magnetic diffusion rate to drop to within a factor of 10 of the
dynamical rate. Thus molecular cloud envelopes and diffuse molecular
cloud tails in cometary structures (e.g., the rho Ophiuchus and Orion
molecular clouds have long cometary tails), should have long magnetic
diffusion times, long lifetimes, and relatively little star formation
\citep{e07}. This is the essential reason why the consumption time of
CO in the galaxy is much longer than the star formation time in each OB
association.  After self-gravity or external pressures compress part of
a cloud, the extinction to the center can exceed several magnitudes.
Then the ionization fraction drops to a new equilibrium regulated by
cosmic rays. Magnetic diffusion becomes rapid in this case. As a
result, rapid star formation requires a threshold cloud column density
of $\sim4$ magnitudes of visual extinction in all directions to
background starlight \citep[8 mag. through the
cloud:][]{mckee89,johnstone04,heiderman10,lada09,lada10}, primarily to
get the magnetic field out of the contracting gas.

There could be a threshold density for star formation as well as a
threshold column density. The distinction between the two may be hard
to determine for solar neighborhood-type clouds.  A threshold density
might arise when the scaling of ionization fraction with density
changes from $n^{-1/2}$ to n$^{-1}$ or faster as a result of changes in
the dominant recombining species (Elmegreen 1978). It could also arise
when atoms and molecules deplete onto grains faster than the dynamical
time, removing key components of the gas-phase ion chemistry, or when
tiny grains grow or get destroyed without reformation, removing an
important component of the total ion collision cross section that
normally resists magnetic diffusion (Elmegreen 2007).  All of these
processes speed up magnetic diffusion and they all seem to operate at a
density between $10^5$ cm$^{-3}$ and $10^6$ cm$^{-3}$ for the solar
neighborhood.  Thus we consider this a threshold density, $\rho_{\rm
thres}$, where the magnetic diffusion rate relative to the dynamical
rate begins to change significantly. Threshold densities have been
observed by \cite{lada97,gao04,wu05,evans08,wu10,lada09,lada10} and
others.

The column density constraint for thermal and radiative shielding,
which gets the ionization fraction into the cosmic-ray dominated
regime, and the density threshold for a suddenly heightened magnetic
diffusion rate, introduce a third component to the theory of star
formation and the IMF discussed in this paper. Although these are
physically and geometrically different conditions, we represent them as
one condition here and simply require the density to exceed a threshold
for our cloud models. In giant molecular clouds, the column density
exceeds the shielding threshold on average before the local density
exceeds the threshold for rapid diffusion. Then the local density
threshold is most relevant for star formation. In other regions where
the density is high and the column density is not, such as supernova
shells, the condition for star formation is the opposite at first
(i.e., before shell gravitational instabilities and local collapse).

A fourth component of the IMF is the fragmentation of cloud structures
during the final stages, when the gas collapses, stars compete for
mass, disks form, interact and shed tidal debris, stars and Brown
Dwarfs get ejected from dense cores, and so on, all shown in numerical
simulations. Fragmentation during and after collapse involves different
physical processes than turbulent fragmentation on GMC scales.
Turbulent fragmentation produces power law mass functions (even with
log-normal density PDFs), but collapse fragmentation can produce
something else. This core fragmentation mass function (CFMF) should be
an important component of the IMF, but it is rarely discussed and it
has not been measured systematically for star-forming regions.
\cite{e00} considered a CFMF that is uniform in $\log M$ and suggested
that the IMF below the thermal Jeans mass reflects this CFMF
exclusively. \cite{swift08} also noted that core fragmentation can
broaden the IMF, while \cite{goodwin08} showed that core fragmentation
into multiple stars can give a better IMF than a one-to-one
correspondence between core mass and final stellar mass. \cite{shad11}
discuss the CFMF in detail.

We find in the cloud models that without a CFMF, the IMF cannot extend
into the Brown Dwarf regime for realistic cloud Mach numbers and
realistic variations of these Mach numbers with position or average
density in a cloud. This is unlike the situation in \cite{padoan02},
who proposed very high and constant Mach numbers to reach a thermal
Jeans mass in the Brown Dwarf regime.  For constant temperature, the
density has to be higher than average by a factor of $100$ to make the
thermal Jeans mass smaller than average by a factor of 10. This seems
necessary if there is no CFMF to broaden the stellar mass range in the
cores.  Getting that far into the tail of a log-normal density PDF
without having the PDF drop too much is difficult. The log-normal PDF
has a count of volume elements at density $\rho$ per unit log-density
that is $\exp\left(-0.5\left[\ln\left\{\rho/\rho_{\rm
pk}\right\}/D\right]^2 \right)$ where $\rho_{\rm pk}$ is the density at
the peak of the log-normal and $D$ is the dispersion. The average
density in this PDF is $\rho_{\rm ave}= \rho_{\rm pk}
\exp\left(0.5D^2\right)$. This means that the maximum value the PDF can
have at a density of $100\rho_{\rm ave}$ occurs for a dispersion where
$\left(\ln \left[100\right] +0.5D^2\right)/D$ is a minimum. This
minimum is at $D=3.035$ and equals the same value, 3.035. For
$D^2=\ln\left(1+0.25{\cal M}^2\right)$ \citep{pnj97}, the Mach number
has to be ${\cal M}=200$ in this case, which is too high, and the PDF
is down from the peak by $\exp\left(-0.5\times3.035^2\right)=0.01$,
which is too low. Alternatively, we could ask what is the ratio of the
density PDF at $\rho=100\rho_{\rm ave}$ compared to the density PDF at
$\rho=\rho_{\rm ave}$? This ratio should be large to have a relatively
large number of Brown Dwarfs in the \cite{padoan02} model. It has a
maximum of unity at infinite $D$, so we ask what is the ratio at some
reasonable $D$, such as that for Mach number ${\cal M}<100$, which
gives $D<2.79$. There, the ratio is 2.6\%, which is also too small.
Thus, a theory of the IMF based on a log-normal density PDF alone, or
even an extension of the log-normal with a power-law to modestly high
density (see below), does not give enough high density material to make
a significant number of Brown Dwarfs if their mass is close to the
local thermal Jeans mass. We need either a density pdf in the
supersonic regime that extends to extremely high density as a power law
or other slowly varying function, or sub-fragmentation inside each
Jeans mass. Brown Dwarf formation at extreme densities in collapse
models was found by \cite{bonnell08}. Sub-fragmentation can give a {\it
range} of stellar or brown dwarf masses that extends downward for a
much larger factor than what the dispersion in the density PDF can give
alone. This is what we mean by a core fragment mass function. It is
different than a single-valued efficiency of star formation in each
core, which does not help to extend the low-mass stellar range relative
to the IMF peak mass. Only a broad function of efficiencies, which is
the CFMF in our model, can do this.

The shape of the CFMF is not known, but if we consider that cloud
structures denser than $\rho_{\rm thres}$ fragment via turbulence into
a power law mass distribution, and that only structures more massive
than the thermal Jeans mass for the local density have the opportunity
to form stars, then the CFMF is the part of the final IMF below the
average thermal Jeans mass. This means that the shape of the CFMF is
the shape of the IMF below the peak around $0.3\;M_\odot$. Note that
the CFMF can be the same for all cloud structures and the summed IMF
from them will still have the power law mass function of the cloud
structure above the peak \citep{e00}. The CFMF is visible on the scale
of clusters and OB associations only for masses below the peak.
However, in very young clusters where the substructures from turbulent
fragmentation are still visible as subclusters of protostars, the CFMF
may be observed directly as the {\it relative} mass distribution
function for these protostars inside each subcluster. If the CFMF is
universal, then the protostellar mass function inside each subcluster,
scaled to the subcluster mass, should be the same for all subcluster
masses. Here is where competitive accretion and other local processes
involving {\it relative mass} should dominate the IMF.

These hypotheses for star formation and the IMF may be summarized as
follows: (1) stars form in cloud structures that are hidden from
outside radiation by $>4$ mag of visual extinction in all directions,
(2) they form fastest where the density $\rho$ exceeds a threshold
value, $\rho>\rho_{\rm thres}\sim10^5m_{\rm H2}$ cm$^{-3}$, (3) they
form in cloud structures that have a $dN/dM\propto M^{-\alpha}$ mass
function ($\alpha\sim2$ to 2.5, the Salpeter function) for all
densities above the threshold and for a wide range of masses, (4) they
form only in those structures that have masses exceeding the local
thermal Jeans mass, and (5) they form by fragmentation in these
structures, with a universal core fragmentation mass function (CFMF).
The density PDF enters the IMF through the distribution function of
Jeans mass \citep{padoan02}, which is the distribution function of the
lower mass limit to the power law for each density exceeding $\rho_{\rm
thres}$.

Whether individual stars form by competitive accretion, using mass from
all over a cloud \citep{bonnell06}, or by monolithic collapse
\citep{shu77,mckee02} using mass from a pre-existing core, would seem
to depend on evolutionary stage. Early on, when filaments are still
draining onto cores, each core and all of its stars effectively accrete
from relatively far away using gas from the filament and beyond. Later
on, when dense cores may become displaced from their filaments or the
filaments become empty, the core evolution is more monolithic. In
addition, for any evolutionary stage, the most centrally condensed
cores will produce the most monolithic-like collapse, i.e., without
severe sub-fragmentation \citep{peters10}. Still, central condensation
is also a result of age: rapid accretion for young cores adds turbulent
energy and mixes the core material, preventing strong central
condensations from starting. All of these processes should happen over
a range of time and spatial scales during star formation, so both
competitive accretion in loose sub-clusters and monolithic collapse
inside centrally concentrated cores should happen simultaneously in a
large cloud complex.

With this model, the total IMF that results from star formation in a
cloud is the integral of the turbulent fragmentation power law over all
the cloud parts denser than $\rho_{\rm thres}$, with lower limits to
each turbulent fragmentation mass equal to the thermal Jeans mass at
the local density.  We write this as
\begin{equation}
P_{\rm IMF}\left(M_{\rm star}\right) = \int_{\rho_{\rm thres}}^{\infty}
M_{\rm gas}^{-2.35} H\left[M_{\rm gas}-M_{\rm J}\left(\rho\right)
\right] P_{\rm PDF,total}\left(\ln\rho\right) d\rho,\label{imf}
\end{equation}
where, P$_{\rm PDF,total}$ is the probability distribution function for
density in the whole cloud, measured per unit $\ln(\rho)$. The mass
integral includes a density multiplied by this probability, $\rho
P_{\rm PDF,total}d\ln\rho$, and $\rho d\ln\rho$ is replaced by $d\rho$.
The other terms are as follows: $P_{\rm IMF}$ is the IMF, i.e., the
probability distribution function of forming a star or Brown Dwarf with
a mass between $M_{\rm star}$ and $M_{\rm star}+dM_{\rm star}$; $H$ is
the Heaviside Function, equal to 0 for negative argument and 1 for
positive argument, and  $M_{\rm J}\left(\rho\right)$ is the thermal
Jeans mass, which is a function of local density, $\propto\rho^{-1/2}$,
assuming a constant thermal temperature for the present paper.

The stellar mass on the left of equation (\ref{imf}), $M_{\rm star}$,
is related to the mass of a gas fragment in the integral, $M_{\rm
gas}$, by the CFMF probability distribution function, $P_{\rm
CFMF}(f)df$ for $f=M_{\rm star}/M_{\rm gas}$. We could therefore
evaluate equation (\ref{imf}) by substituting $M_{\rm star}/f$ for
$M_{\rm gas}$ in the integral, multiplying the result by $P_{\rm
CFMF}(f)$, and integrating over $f$ from 0 to 1. When $M_{\rm
star}<M_{\rm J}$, the Heaviside function of $(M_{\rm star}/f-M_{\rm
J})$ cuts the integration limits for $f$ to 0 and $M_{\rm star}/M_{\rm
J}$, making the IMF depend on the shape of $P_{\rm CFMF}(f)$. When
$M_{\rm star}>M_{\rm J}$ the integration limits for $f$ are 0 to 1 and
the integral over $f$ is independent of $M_{\rm star}$, leaving the IMF
with the $M_{\rm star}^{-2.35}$ dependence from the clumps. Other
analytical expressions involving the CFMF are in \cite{e00} and
\cite{shad11}.

To evaluate the IMF numerically, we consider that the relative
proportion of the incidents of $f$ is $P_{\rm CFMF}(f)$, so the
distribution of $M_{\rm star}$ for each $M_{\rm gas}$ comes from the
equation
\begin{equation}
{{\int_{M_{\rm min}/M_{\rm gas}}^{M_{\rm star}/M_{\rm gas}} P_{\rm
CFMF}(f) df} \over {\int_{M_{\rm min}/M_{\rm gas}}^{1} P_{\rm CFMF}(f)
df} } = X,\label{eq:x}\end{equation} where X is a number uniformly
distributed between 0 at $M_{\rm star}=M_{\rm min}$ and 1 at $M_{\rm
star}=M_{\rm gas}$ ($X$ can be a random number or a sequence of
regularly spaced numbers in this interval). In the lower limit of the
integral, $M_{\rm min}$ is taken equal to $0.01$. To evaluate the IMF,
we determine $P_{\rm PDF,total}(y)$ as discussed above, normalize
$M_{\rm J}$ to unity at $y=y_{\rm thres}$, and choose $P_{\rm
CFMF}(f)=1/f$ to make the IMF flat below $M_J$.

\begin{figure}\epsscale{.65}
\plotone{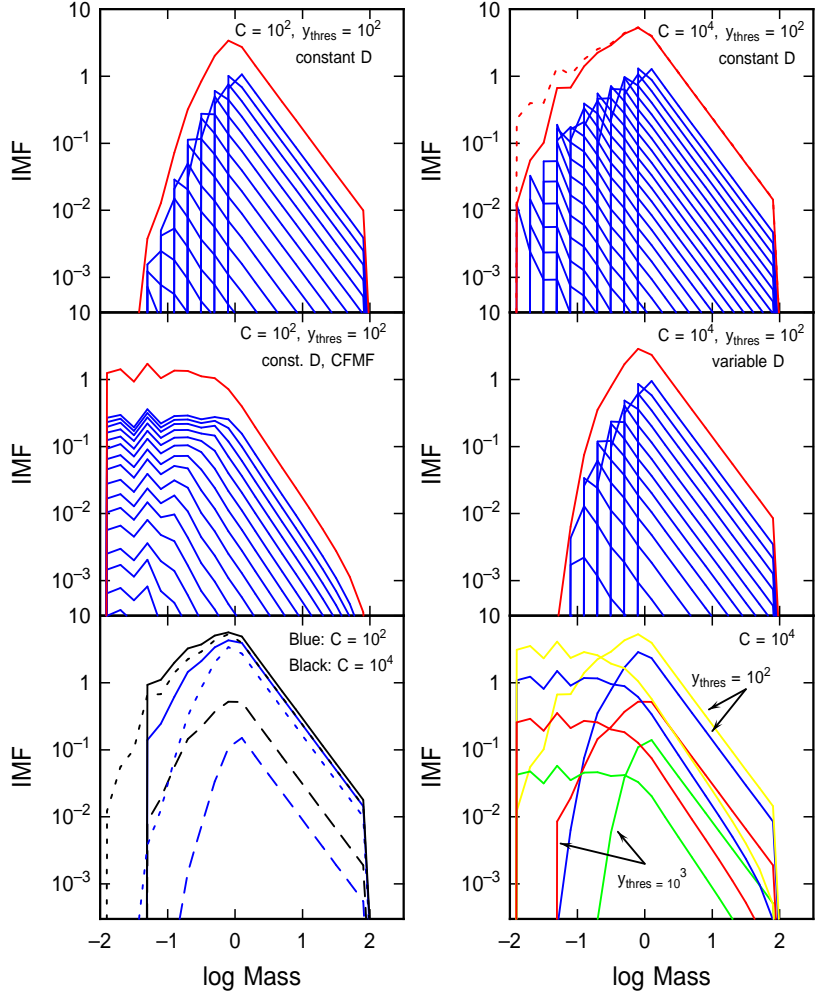} \caption{Stellar IMFs for a model with a Salpeter
slope for cloud cores more massive than the local thermal Jeans mass,
and either a direct correspondence between stellar and core masses
(peaked curves) or a core fragmentation mass function that has a
uniform probability for the star/core mass fraction per unit log $M$
(curves with a flat part at low $M$). Other assumptions in the IMF are
a threshold density $y_{\rm thres}$ and no star formation in cores less
massive than the Jeans mass.  Red curves in the top four panels are
summed IMFs for the whole cloud, and blue curves are components of this
IMF for different relative local densities $y$. ${\cal C}$ is the cloud
concentration parameter and $D$ is the dispersion in the log-normal
part of the convolution-PDF. In the lower left, solid-line curves have
a constant Mach number ${\cal M}=20$ and $y_{\rm thres}=10^3$; dotted
curves have constant ${\cal M}=5$ and $y_{\rm thres}=10^2$; dashed
curves have constant ${\cal M}=5$ and $y_{\rm thres}=10^3$. In the
lower right, ${\cal C}=10^4$ for all cases; yellow curves have $y_{\rm
thres}=10^2$ and constant Mach number ${\cal M}=5$; blue curves have
$y_{\rm thres}=10^2$ and variable Mach number with edge value ${\cal
M}_{\rm edge}=5$; red curves have $y_{\rm thres}=10^3$ and ${\cal
M}=5$, while green curves have $y_{\rm thres}=10^3$ and variable Mach
number with ${\cal M}_{\rm edge}=5$. Each color in the lower right has
one curve with no CFMF and another curve with a
CFMF.}\label{convolution_imf}\end{figure}

Figure \ref{convolution_imf} shows sample IMFs generated from equations
(\ref{imf}) and (\ref{eq:x}). In the top four panels, the summed IMF
for the whole cloud is shown as a solid red curve (at the top of all
the other curves), and the IMFs for each relevant $y$ value, with
logarithmic spacings of $y$, are shown as blue curves (recall $y$
measures the local density, including turbulent compression:
$y=\rho/\rho_{\rm edge}$). A relevant $y$ value is one that exceeds the
threshold density for star formation. In the top two panels and
middle-right panel, there is no clump fragmentation mass function
($P_{\rm CFMF}(f)=$ a delta function, i.e., $f$ is constant). In the
middle-left panel, there is a CFMF, $P_{\rm CFMF}(f)=$constant in equal
intervals of $\ln f$.  When there is no CFMF, each blue curve consists
of a power law at masses larger than the Jeans mass for that $y$ value,
and a sudden drop at lower mass. This illustrates our assumption that
stars form only with masses above the local Jeans mass, with no spread
from a fragmentation mass function in these cases. Also in these three
panels, increasing $y$ corresponds to a decreasing local Jeans mass,
and therefore a decreasing peak mass in the blue curve. When there is a
CFMF, in the middle-left panel, each blue curve for separate $y$ again
has a power-law decrease above the local Jeans mass, but now there is a
spread in stellar mass below the local Jeans mass from the CFMF.
Because we assumed in this case $P_{\rm CFMF}(f)=$constant in equal
intervals of $\ln f$, the IMF below each local Jeans mass is also
constant for equal intervals of $\log M_{\rm star}$, i.e., the blue
curves are flat.

The summed IMF in each of these four cases has the same general shape
as the local IMFs. When there is no CFMF, the summed IMF has a power
law above the largest value of the local Jeans mass and a curving
downward trend below this largest Jeans mass. The curve downward is
from the decreasing total mass of cloud gas at higher and higher
densities (higher $y$).  When there is a CFMF, the sum at intermediate
to high mass is still a power law, as the CFMF does not reveal itself
for masses larger than the largest local Jeans mass. At masses below
this largest Jeans mass, the summed IMF is the CFMF, i.e., $\propto
P_{\rm CFMF}$.  The largest Jeans mass occurs at the smallest density
exceeding the threshold for star formation, namely at the threshold
density itself. For all cases, we set the local Jeans mass to be
$M_{\rm Jeans}=\left(y_{\rm thres}/y\right)^{1/2}$, so at the threshold
the value is unity. This is not supposed to represent a value in solar
masses, as the physical dimensions are not considered here. For a
discussion of physical processes that could influence the mass at the
peak of the IMF, see \cite{ekw} and \cite{bate09}.

Again considering the top two panels of Figure \ref{convolution_imf}
and the middle-right panel, all without the CFMF, the IMF below the
largest Jeans mass depends on the cloud density concentration ${\cal
C}$.  As ${\cal C}$ increases, the IMF spreads to lower mass. The
dotted red line in the top right panel has ${\cal C}=10^6$ with all
else the same. The trend to fill in the IMF toward lower mass stars
continues for this higher concentration.  In all cases, the IMF peak is
at the largest Jeans mass.

The top and middle right-hand panels illustrate the difference that a
constant or variable Mach number makes. The top right panel has a
constant Mach number (${\cal M}=5$) and therefore constant dispersion
$D$ in the log-normal part of the PDF, and the middle right panel has a
Mach number that decreases inside the cloud according to equation
(\ref{eq:mach}) with ${\cal M}_{\rm edge}=5$.  When the Mach number
decreases, the total PDF drops down at large density (Figure
\ref{convolution} dotted blue lines), and so the IMF drops fast at low
mass. Physically, this is because there is relatively little
compression from turbulence in the inner regions of the cloud, and so
little mass at high enough density to make low-mass stars greater than
their local Jeans mass.

The bottom two panels of Figure \ref{convolution_imf} show many
total-cloud IMFs to illustrate the dependence of the IMF on various
parameters. In the lower left, blue and black curves are for ${\cal
C}=10^2$ and $10^4$ respectively. The higher ${\cal C}$ curves always
extend to lower mass for the same line type, as explained two
paragraphs above. The solid curves are for a Mach number equal to a
constant value of 20 and a threshold normalized density of $y_{\rm
thres}=10^3$. The dotted curves are for a constant Mach number of 5 and
$y_{\rm thres}=10^2$. The dashed curves are for constant ${\cal M}=5$
and $y_{\rm thres}=10^3$. In all cases considered in this paper, a
maximum value of $y_{\rm max}=10^6$ is assumed. Thus the minimum
stellar mass that can form without a CFMF occurs at $\log
M=\left(y_{\rm thres}/y_{\rm max}\right)^{1/2}$. When $y_{\rm
thres}=10^3$ in the solid and dashed curves, the minimum stellar mass
is $10^{-3/2}=0.031$. Plotting in all cases is by histogram with $\log
M$ intervals unrelated to the PDF calculation intervals for $\ln y$.
This explains the bin-to-bin irregularities in the plotted IMFs and the
slight offset in the minimum mass bin for the $y_{\rm thres}=10^3$ case
from the expected value of 0.031.  In summary, the comparison in the
lower left panel shows that higher $y_{\rm thres}$ produces lower
numbers of stars because there is less gas mass exceeding the threshold
density for star formation. It also shows that lower Mach numbers
produce slightly fewer low mass stars (dotted curves compared to solid
curves) because there is less turbulence compression at lower Mach
numbers. This effect is minor because the Mach number enters the
dispersion in the log-normal part of the PDF only weakly.

The lower right panel of Figure \ref{convolution_imf} compares
different $y_{\rm thres}$, different ${\cal M}$ variability, and
different CFMF's, all for the same cloud concentration factor of ${\cal
C}=10^4$.  The four curves that have a peak are labeled with their
$y_{\rm thres}$ values, $10^2$ for the two higher curves and $10^3$ for
the two lower curves. The topmost in these pairs (yellow and red
curves) have a constant Mach number ${\cal M}=5$, and the lower curves
in these pairs (blue and green) have variable Mach numbers with ${\cal
M}_{\rm edge}=5$. When $y_{\rm thres}$ is low, the variable Mach number
does not matter much because star formation is easy and occurs in most
gas, even with little ram-pressure compression.  Thus the power-law
parts of the blue and yellow curves are similar. When $y_{\rm thres}$
is high, the variability of the Mach number matters more, i.e., the
power-law parts of the red and green curves differ.  This difference is
because star formation at large $y_{\rm thresh}$ is confined to the
most highly compressed regions.

The four other curves in the lower right panel of Figure
\ref{convolution_imf}, which have flat IMFs at low mass, use the CFMF
mentioned above, i.e., $P_{\rm CFMF}(f)=$constant in equal intervals of
$\ln f$. The color coding is the same as for the peaked curves: the top
two have $y_{\rm thres}=10^2$ and the bottom two have  $y_{\rm
thres}=10^3$, while the topmost of each has a constant Mach number and
the bottom curve of each has a variable Mach number. The trends are the
same as for the peaked curves and for other panels in this figure:
higher $y_{\rm thres}$ produce fewer stars and lower IMF curves;
variable Mach numbers produce less high density gas and fewer low mass
stars relative to high mass stars, and a CFMF produces a spread in the
stellar mass for each cloud core mass, which shows up as a spread in
the IMF below the largest Jeans mass.

The lower right panel also shows how insensitive the IMF is to cloud
parameters in the case where there is a CFMF. All that varies is the
height of the summed IMF, which means the overall efficiency of star
formation (as discussed also for Figures \ref{convolution} and
\ref{convolution_eps}). When there is a CFMF, the IMF does not
noticeably depend on the cloud concentration factor (compare the middle
left panel to the lower right panel of Fig. \ref{convolution_imf}), the
threshold density, or the Mach number. All of these details are hidden
by the CFMF, which dominates the IMF below the peak. Without the CFMF,
the low mass part of the IMF depends on ${\cal C}$, $y_{\rm thres}$,
and the value and variability of ${\cal M}$.

\section{Conclusions}

The density PDFs of molecular clouds have extended tails from the
concentration of mass near the cloud center or in regular structures
like self-gravitating filaments and cores.  The log-normal PDF from
turbulence should be generated only locally where the average density
and Mach number are relatively uniform. Thus a reasonable model for the
total PDF is a convolution of the local log-normal with the density and
velocity structures in the cloud. Several examples of this
convolution-PDF were generated here, showing the effect of the density
concentration factor, ${\cal C}$, in converting a log-normal tail into
a power law tail for clouds with power-law radial density profiles and
constant Mach number. Variable Mach numbers present a different case as
the local PDF can become narrow when the average density is high, and
then the total PDF falls toward high density nearly as fast as in the
case without a density concentration. Power-law PDF tails increase the
mass fraction of the gas above a threshold density for star formation,
thereby increasing the absolute star formation rate, the star formation
rate per unit gas mass, and the final efficiency.  These increases take
place mostly in the cloud center, where the high average density is
compressed to even higher values by turbulence. As a result, strongly
gravitating clouds form stars much faster than weakly gravitating
clouds, and bound clusters can form quickly in the core of a centrally
concentrated cloud, even when the average efficiency of star formation
throughout the cloud is low.

A reasonable IMF model produces a power-law stellar mass function from
cloud structure for intermediate-to-high mass stars above the average
thermal Jeans mass, and a decreasing stellar mass function below the
IMF peak from the decreasing total mass of gas at high density. In all
cases, the core mass that produces one or more stars exceeds the local
Jeans mass. We follow this model here, which was originally expressed
by \cite{padoan02}, with two additional assumptions: a minimum density
threshold for star formation and a core fragmentation mass function.
The threshold density limits the cloud mass and the regions in a cloud
where stars can form, placing most star formation in a cloud core where
the net efficiency and rate are large, and limiting the thermal Jeans
mass to a maximum value at the threshold density (assuming a constant
temperature).  The core fragmentation mass function produces a spread
of stellar masses for each core mass. The CFMF assumed here is a
function of the relative masses for stars and cores, not an absolute
mass function for the stars. Thus more massive cores produce more
massive stars with the same relative mass distribution. The result of
such a relative CFMF is that the stellar mass distribution follows the
core mass distribution, i.e., the cloud structure mass distribution,
for all core masses above the thermal Jeans mass, and it extends down
to low stellar or Brown Dwarf masses below the core thermal Jeans mass
with a shape that is identical to the CFMF.  The CFMF may be observed
as a relative mass distribution function for stars inside the
sub-condensations of molecular clouds or simulations. We predict that,
scaled to the sub-condensation mass, the stellar mass functions inside
those sub-condensations will all be about the same. The basic model for
this was proposed in Elmegreen (2000).

This model of cloud evolution proposes that the structures which form
stars are mostly in existence before star formation begins. They may
not be in round cores, but, whatever their shapes, their mass
distribution functions are in place before star formation, and their
locations and relative masses in the overall cloud are in place too.
Two advantages of this model over those with more extended periods of
structure formation and re-formation (i.e., for more than a few
dynamical times) are that star formation can be rapid in the present
model, and an IMF determined early should not be strongly affected by
stellar feedback that comes later.

Helpful comments by E. V\'azquez-Semadeni are appreciated.

{}

\clearpage

\clearpage

\clearpage
\end{document}